\pdfoutput=1
\documentclass{article}

\PassOptionsToPackage{numbers, compress}{natbib}

\usepackage[preprint]{neurips_data_2023}

\usepackage[utf8]{inputenc} 
\usepackage[T1]{fontenc}    
\usepackage{url}            
\usepackage{booktabs}       
\usepackage{amsfonts}       
\usepackage{nicefrac}       
\usepackage{microtype}      
\usepackage{xcolor}         
\usepackage{graphicx}
\usepackage{hyperref}
\hypersetup{colorlinks,allcolors=black}
\usepackage{natbib}
\usepackage{cleveref}
\usepackage{listings}
\usepackage{graphicx}
\crefname{figure}{Figure}{Figure}
\crefname{table}{Table}{Table}
\crefname{section}{Section}{Section}
\crefname{appendix}{Appendix}{Appendix}
\title{LSCD: A Large-Scale Screen Content Dataset for Video Compression}

\author{
    \begin{tabular}[t]{ccc}
        \textbf{Yuhao Cheng} & \textbf{Siru Zhang} & \textbf{Yiqiang Yan} \\
        \texttt{chengyh5@lenovo.com} & \texttt{siru.zhang@outlook.com} & \texttt{yanyq@lenovo.com}\\
    \end{tabular}\\
    \begin{tabular}[t]{cc}
        \specialrule{0em}{5pt}{10pt}
        \textbf{Rong Chen} & \textbf{Yun Zhang} \\
        \texttt{ronechen216@gmail.com} & \texttt{zhangyun1202@outlook.com} \\
    \end{tabular}\\
    \begin{tabular}[t]{c}
        \specialrule{0em}{15pt}{0em}
        \textbf{Lenovo Research}\\
    \end{tabular} 
}

\begin{document}

\maketitle
\begin{abstract}
Multimedia compression allows us to watch videos, see pictures and hear sounds within a limited bandwidth, which helps the flourish of the internet. During the past decades, multimedia compression has achieved great success using hand-craft features and systems. With the development of artificial intelligence and video compression, there emerges a lot of research work related to using the neural network on the video compression task to get rid of the complicated system. Not only producing the advanced algorithms, but researchers also spread the compression to different content, such as User Generated Content(UGC). 
With the rapid development of mobile devices, screen content videos become an important part of multimedia data. In contrast, we find community lacks a large-scale dataset for screen content video compression, which impedes the fast development of the corresponding learning-based algorithms.  
In order to fulfill this blank and accelerate the research of this special type of videos, we propose the \textit{Large-scale Screen Content Dataset(LSCD)}\footnote[1]{If you want to get the dataset, codes and trained models, please contact the first author through the printed email or yuhao.cheng@outlook.com.}, which contains 714 source sequences. Meanwhile, we provide the analysis of the proposed dataset to show some features of screen content videos, which will help researchers have a better understanding of how to explore new algorithms. Besides collecting and post-processing the data to organize the dataset, we also provide a benchmark containing the performance of both traditional codec and learning-based methods.

\end{abstract}
\section{Introduction}
Data compression is a critical technology for transmitting and storing data. Especially for multimedia data, the compression technology could reduce the redundancy inside the data and economize resources to use them. 
So, how to efficiently compress multimedia data such as images and videos has always been one of the core technologies in signal processing and computer vision. Before the bloom of the artificial neural network, researchers have proposed many excellent hand-craft methods to compress the images and videos, such as H.264\cite{wiegand2003overview}, H.265\cite{sullivan2012overview} for videos and JPEG\cite{marcellin2000overview} for images. With the development of artificial intelligence, especially the neural networks, there is more and more work proposing using the neural network to compress the images\cite{balle2016end} and videos\cite{lu2019dvc,yilmaz2021end,chen2021nerv}. 

\begin{figure}
    \includegraphics[width=\linewidth]{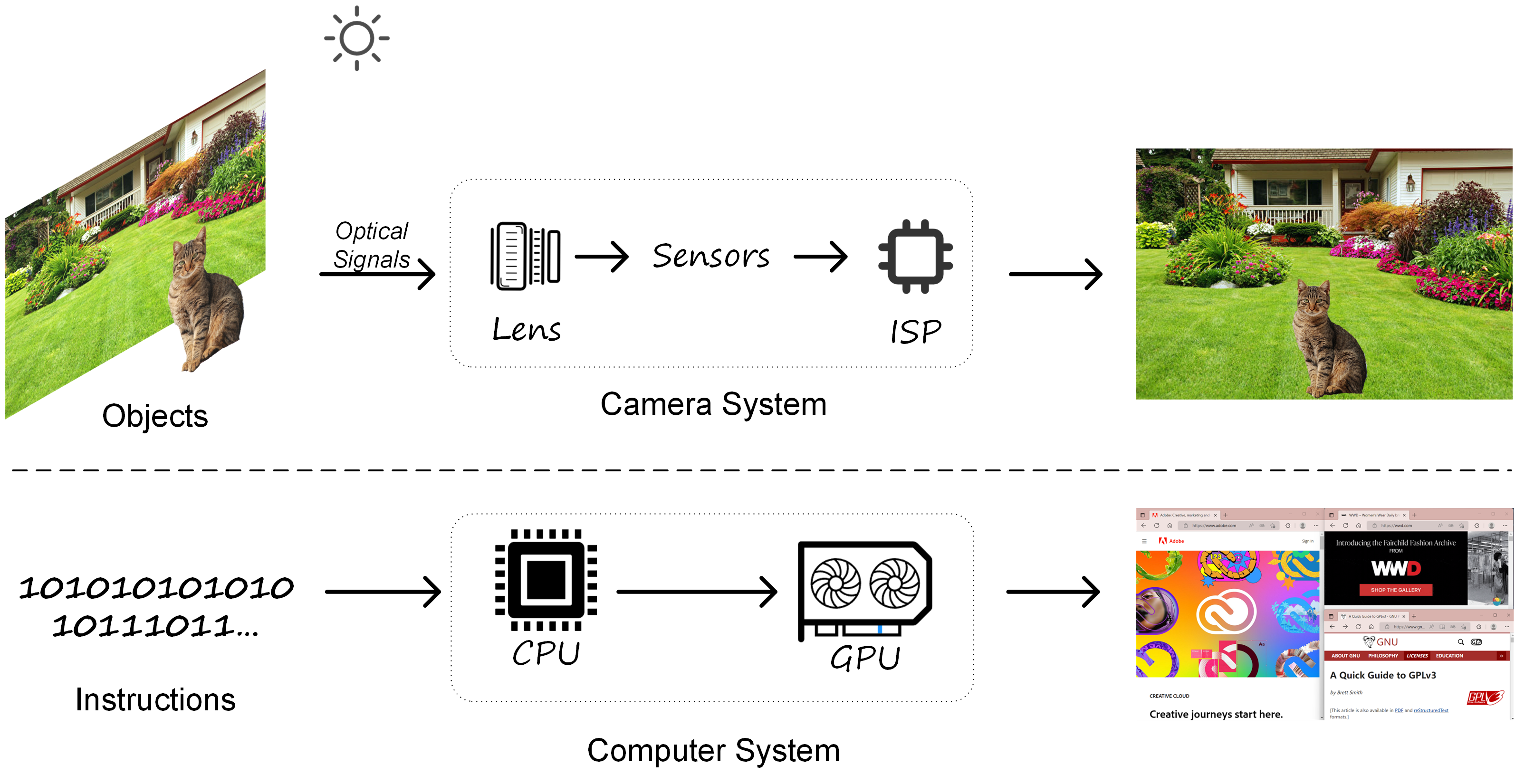}
    \caption{A simple illustration of the difference in how to generate images between natural contents and screen content. The upper one is the ordinary way to get the natural contents through cameras. And the lower one is what happens in the computer when we get the screen contents.
    }
    \label{fig:diff_camera_gpus}
\end{figure}

The information from nature is huge and countless. Obviously, there are lots of redundancy when we want to transmit, store and use these data and the aim of compression methods is to reduce these redundancy without disastrously decreasing the quality of the data. For image data, it has spatial redundancy in itself and the entropy redundancy during the coding process. Compared with image data, video data is more complicated and contains temporal information. As a result of that, compressing the video will reduce the entropy, spatial and temporal redundancy. 
And not surprisingly, at present, most of the algorithms are focused on how to use the neural network to compress the data obtained from the nature. 
Meanwhile, with the rapid development of the mobile and personal devices such as personal computers and mobile phones, the data generated by them, screen content videos, gradually becomes one of the important multimedia data in industry and research. Previously, the screen of devices always only display the videos or images generated by cameras, however, at present, we more and more transmit video from the screens, or in other words, the screens become another "camera" to obtain the multimedia data. 
Moreover, the processes of getting the natural content and screen content videos are different, which is shown in \cref{fig:diff_camera_gpus}. \cref{fig:diff_camera_gpus}'s upper part shows the ordinary way to get the natural content. The camera receives the lights reflected by the objects through the lens, and then the sensors(e.g. CMOS) will convert the lights into digital signals. Finally, the images will be produced by the ISP(Image Signal Processor). \cref{fig:diff_camera_gpus}'s lower part shows what mostly happens in the computers or other end devices. The CPU(Central Processing Unit) receives the instructions and transmits to the rendering part, such as GPU(Graphical Processing Unit). And the GPU will render the images based on these instructions. In summary, for cameras, the images or videos are the reflections of the lights from objects, while for screens, these are directly from the rendering of the GPUs. 
As a result, intuitively, the data from cameras and screens are different, and they may have some unique features. 
So focusing on the screen contents, we propose the first Large-scale Screen Content Dataset(LSCD). 

The contributions are as following:
\begin{itemize}
    \item Produce the first large-scale screen content dataset for video compression. Through collecting the various files and defining the actions related to these files, we build the proposed dataset.
    \item Provide the analysis of the screen content videos. Basing on the proposed data, we leverage the common used video descriptors to analyze the proposed dataset. In this way, we show the distribution of the proposed dataset and some features related to the screen content videos. 
    \item Build a benchmark of video compression methods on the proposed dataset. Not only producing the dataset, we also do lots of experiments to build a benchmark of the proposed dataset to see the potential research points of this task.
\end{itemize}
\section{Related Work}
\textbf{Compression Algorithms.} How to compress the data is always one of the core technologies and research topics. The compression technologies can be divided into two main parts based on whether we can recover the data with or without loss. Specifically, \textit{lossless} methods guarantee that the compressed data could be recovered to the exactly same data as the original one, while \textit{lossy} methods aim to use fewer bits to represent the data with the tolerable difference between the original data. As for the multimedia data, the compression algorithms are mostly the lossy ones, such as JPEG\cite{marcellin2000overview} for images and H.264\cite{wiegand2003overview}, H.265\cite{sullivan2012overview} for videos. 
Although traditional methods achieve excellent performance, it has become a complex system with an elaborated design of each module and the combination of different modules, making it difficult to adjust it on other data. Moreover, with the development of artificial neural networks and deep learning, many AI-based methods have outstanding performance in image and video compression. Comparing with the traditional methods, which use elaborate hand-craft features and modules, the deep learning one uses abundant data to train a model which could compress different types of images and videos. 
As a result of the good generalization of neural networks, we don't need to manually adjust the models to fit the different data in most situations.
There are lots of deep learning methods, such as \cite{lu2019dvc,yilmaz2021end,chen2021nerv,hu2021fvc,wang2022neural,rippel2019learned,agustsson2020scale}. Nevertheless, if the aimed data goes to another domain which is far from the training data, we still need the corresponding data to get a new model with existing or new algorithms and screen content videos lie in the another domain compared with the natural videos.  
And our dataset provides a new area both for the traditional codecs and deep-learning methods to explore the latest algorithms and validate them.

\textbf{Datasets.} Data is one of the vital factors accelerating the development of data-driven artificial intelligence. There are lots of famous datasets that help the research community, for examples, ImageNet\cite{deng2009imagenet} helps the classification task and shows the potential power of the neural network; COCO\cite{lin2014microsoft} makes the detection, pose estimation, visual question answering, etc. have a significant step forward; LAION-5B\cite{schuhmann2022laion} helps the multimodality community could research on the large-scale data. Previously, we also had some datasets related to video and image compression. Kodak\cite{kodak} is the classical dataset for image compression; MCL-JCV\cite{wang2016mcl} produces a dataset for JND-based video quality assessment; BVI-DVC\cite{ma2021bvi} provides the lots of natural videos to help researchers in video compression; SCVD\cite{cheng2020screen} is a dataset aiming at the video quality of the screen content. Compared with these previous datasets, our dataset focuses on the screen content videos and collects the data close to the actual situation when users transmit the screen videos.  Although our proposed dataset focuses on the video data, these frames of dataset can be used for image compression.
\section{LSCD Main Features}
Compared with the previous dataset, our proposed LSCD dataset has following features: screen content oriented, lossless and large scale. And we think these features will accelerate the research on the multimedia compression community. The comparison between the proposed and previous datasets is shown in \cref{tab:dataset_comparison}. 
\begin{table*}[t]
  \centering
  \resizebox{\linewidth}{!}{
  \begin{tabular}{c|ccccccccc}
  \hline
                    & Kodak\cite{kodak}   & DIV2K\cite{agustsson2017ntire}  & Vimeo-90K\cite{xue2019video}  & UVG\cite{mercat2020uvg}    & MCL-JCV\cite{wang2016mcl} & BVI-DVC\cite{ma2021bvi} & SCVD\cite{cheng2020screen}    & Ours    \\ \hline
  Image or Video?   & Image   & Image  & Image  & Video  & Video   & Video   & Video   & Video   \\
  Source Seq Number & 24      & 1000   & 89800  & 16     & 30      & 200     & 16      & 714     \\
  Max Resolution    & 768x512 & 2040p  & 256p   & 2160p  & 1080p   & 2160p   & 1080p   & 1080p   \\
  Original Format   & PNG     & PNG    & PNG    & YUV    & YUV     & YUV     & YUV     & BMP/PNG     \\
  Bit Depth         & 8       & 8      & 8      & 8/10      & 8       & 10      & 8       & 8       \\
  Content Type      & Nature  & Nature & Nature & Nature & Nature  & Nature  & Desktop & Desktop \\ \hline
  \end{tabular}
  }
  \caption{The comparison between different multimedia compression datasets. The \textit{Source Seq Number} means the number of videos which is used to generate the rest of the data. For example, BVI-DVC dataset has 200 4K videos and they resize the original videos to get the rest part of the dataset.}
  \label{tab:dataset_comparison}
\end{table*}

\textbf{Screen Content Oriented.} Firstly, the proposed dataset focuses on screen content data. Most previous video compression datasets focus on natural content, and we do not have a large-scale dataset containing screen content data for video compression. 
As we have mentioned before, the screen content videos lie in a different domain from that of the natural content. And because of the difference in how images are generated, the screen videos have more high-frequency parts, for example, more edges and characters. Meanwhile, we still have some natural videos, however, they are played by some players on the computer to simulate real situations. In this way, we convert these natural videos to screen content videos. 

\textbf{Lossless Data.} Secondly, the proposed dataset is lossless. Not like the other computer vision tasks, the video compression is so sensitive to whether the original data is lossless or lossy. If the original training content videos are compressed by the third-party, the proposed methods are making a simulation of these compression algorithms. However, as for the computer generated content, we could avoid this problem by getting the images directly from the graphic card. 
LSCD contains the BMPs of each frame which are honestly storing the full information of the screens. 
The advantages of this are that in the future, we can manipulate these to produce the data we need and we also could directly explore the intrinsic features of these videos.

\textbf{Large Scale.} Thirdly, the dataset is large-scale. It goes into two parts, one is the number of the video, and the other one is the length of the video. The previous screen content video compression dataset does not contain so many videos. And meanwhile, in the previous dataset, each video does not have so many frames. We use so many frames because we find that we want to simulate the real scene in which we use video compression in the streamline. The previous dataset always contains a short period of the video, in that case, the method may not have the chance to explore the long-term temporal information. And we also can break the long video into short ones easily to follow the previous settings. 
\section{Collection Methodology}
For collecting the data, we breakdown the problem into its basic units, one is the file and the other one is the action. So the problem turns to how to collect the files and how to define the actions related to these files and softwares. 
And the \cref{sec:assembly} will introduce the steps in details following the clue of this idea. Meanwhile, the protection of the private information during collecting is also an important part. So \cref{sec:privacy} will explain what we have done in this aspect.
\subsection{Dataset Assembly Pipeline}
\label{sec:assembly}

\textbf{Recording Tool Implementation.} Firstly, we build a tool to record lossless desktop videos. Although we have lots of softwares to record the screen, we find that we can not have full control of recording the screen if we use third-party softwares. Meanwhile, the current recording softwares, in order to reduce the size of the recording file, will compress the file. However, we need the uncompressed videos. So we implement a tool to directly get the BMPs by using the GDI without any compression. 

\textbf{Content materials collections.} In the daily use of the screen, documents are a huge part of all materials. And different materials will have different graphical structures. For example, MS Word may contain lots of characters, while Power Point may have more pictures. So we collect many different kinds of documents to try our best to cover the different graphical structures.

\textbf{Defining actions.} After we have collected the materials, we need to define actions that could reflect the real situation when users use these materials. 
To be honest, the designing of actions is more complex than directly recording the videos. On the one hand, we need to design these actions as real as possible. On the other hand, actions should not be so well designed, which are not similar to the daily use. And in order to solve the problem, we first design three levels of complexity, and then on different levels, we use the combination of these materials and motion to build each data sample. Moreover, we additionally build a \textit{supp} part which contains some ordinary scenarios but is hard for the current video codecs. 

\textbf{Collecting.} Based on the definitions of each data sample, we asked some volunteers to record the data samples. During this phase, we tell them that their screen will be recorded and ask them to use the documents which we have collected before. Notably, we do not ask them to use the same computers, and we just ask them to set the same resolution and the FPS. We think there are slight differences between different computers, monitors and etc., so we use multiple computers to record the videos. In this way, we want to import some noise related to the devices, which will help our dataset to cover a wider range of the data distribution. 

\textbf{Post-processing.} After getting raw data, we post-process these data. 
The post-processing mainly contains two parts. One is to check whether the data fits the requirement, such as the resolution, FPS, and so on. The other is to fix some bad cases, such as the wrong storage. Because we record the video for a long time, sometimes there will be a memory wrong which causes the BMP will not be stored correctly. Fortunately, we found that we could solve this problem by re-collecting the corrupted data. 
Meanwhile, we make the lossless data to reduce the need for the storage of the dataset, specifically, we change the BMPs to PNGs. 

By following these steps, and sometimes we need to repeat some of them to make sure that the original data does not have any problems. After that, we get the LSCD dataset.
\subsection{Privacy During Collection}
\label{sec:privacy}
As desktops are one of the private digital space of people, we should not leak any privacy information during the dataset collection process. And in the collection, we use the following methods to protect the information. 

\textbf{Publicly available docs.} All of the materials used in building the dataset are publicly available on the internet. Instead of using personal documents, we use the files from the internet. Meanwhile, we choose the files close to the actual usage, although there are so many template files on the internet. For example, we have so many power points templates, and most of them have so many blanks for users to fulfill. And in actual situation, we do not need these blanks. Through this way, we protect the privacy from materials.

\textbf{Collectors know the data collection process.} We only collect the data when the collectors know that and do not use the back thread to collect the data. 
And before recording the data, we will tell them which kinds of private information are so sensitive that shouldn't be shown during the process of recording the data. Except these sensitive information, the other information is decided by the volunteers themselves. 
\section{Dataset Composition}
We firstly define four levels of actions and collect each of them. The reason why we define these different scenes is that we want to reflect different usage of devices. The scenes' definitions are shown in the following:
\begin{itemize}
  \item \textbf{Plain.} This level is to simulate the simple situation when users use the computer in their daily life, such as opening a document to read, watching a video, standing in the desktop and so on. We collect 30 videos under this level.
  \item \textbf{Medial.} This level is to simulate that users use the computer with a little bit complex, such as installing the softwares, manipulate the documents and so on. In this level, it has the videos including the more complex screen contents and motion. We collect 281 videos under this level. 
  \item \textbf{Complex.} This level is more like the scenarios when we need to deal with multiple tasks synchronously. So in this level, the data will contains the examples of opening many windows and so on. We collect 293 videos under this level.
  \item \textbf{Supp.} This level contains some complicated cases that it may be difficult for both traditional and deep learning methods. We collect 110 videos under this level.  
\end{itemize}
The dataset is split into two parts: training and testing. For the traditional codec, we may not need the data to train, while for data-driven methods, we need a part to train the model. And for building the test set, we randomly select $20\%$ of each scene. However, we do not choose the data points in the \textbf{Supp} because we want to make the dataset more challenging for the model. 
So the composition of our proposed dataset can be summarized in the following:
\begin{itemize}
  \item \textit{Training Set.} The function of this part is training or fine-tuning the future deep models. And followed by the common standard, we randomly select $80\%$ of \textbf{plain}, \textbf{media} and \textbf{complex} respectively to form the training set. This part contains 234 complex videos, 224 medial videos and 6 plain videos which has 1326725 frames in total. 
  \item \textit{Testing Set.} The function of this part is testing the deep learning models or traditional codecs. And we use the remain of the \textbf{plain}, \textbf{media} and \textbf{complex} to make up the testing set. This set contains 59 complex videos, 57 medial videos and 6 plain videos which has 358309 frames in total. 
  \item \textit{Additional Testing Set.} This part contains the \textbf{supp} part and is used to test the algorithms in much more complicated situations. This set contains 110 videos which has 35639 frames in total. 
\end{itemize}
\begin{figure}
    \includegraphics[width=\linewidth]{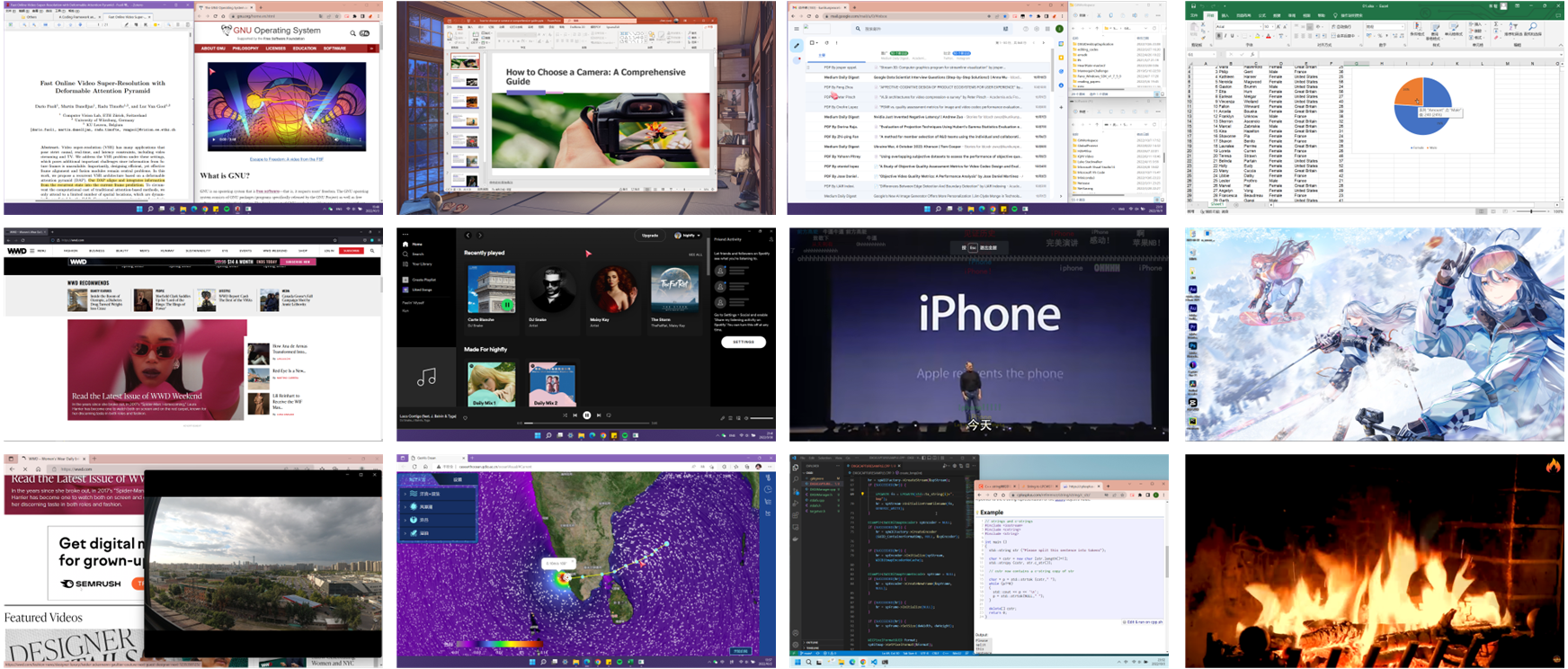}
    \caption{Overview of the LSCD. The figure shows some representative data in our proposed dataset(Better in color). And we can not show all the figures of the whole dataset, there are more pictures in the supplemental material.}
    \label{fig:dataset_preview}
\end{figure}
\section{The Analysis Related to LSCD}
During making the dataset, we want to try our best to make the proposed dataset could reflect the feature of screen content videos, meanwhile, not only fitting some specific situations. So in this part, we will use two parts of analysis to show that our dataset has the ability to help researchers in the future work. The first part is analysis from the subjective aspect, and the other part is from the objective aspect.
\subsection{Subjective Analysis}
In this part, we want to show the content of the proposed dataset. 
From \cref{fig:dataset_preview}, we can know that LSCD contains scenarios which are usually met in the daily work, such as processing the kinds of documents, scanning the webpages, watching videos, and so on. Meanwhile, all of these videos are 1080x1920(height x width) resolution and 25 FPS, which are the most common settings when we use the screen. Because we have so many different videos, we can not list all of them here, and you can scan them by downloading our dataset. In a word, we try our best to collect as many as videos to reflect the actual situations of daily use. 
\subsection{Objective Analysis}
\begin{figure}
    \centering
    \includegraphics[width=0.8\linewidth]{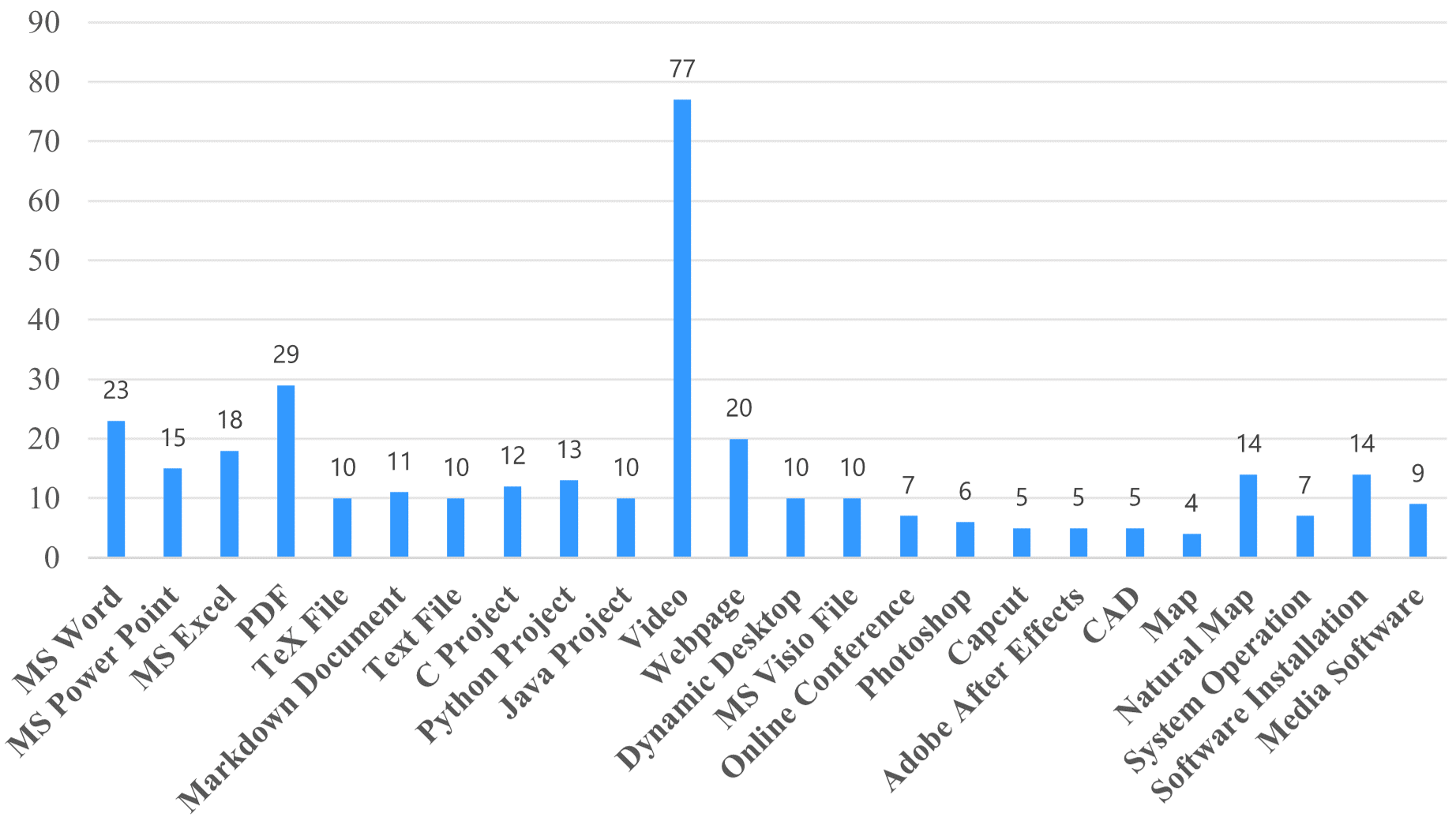}
    \caption{The statistical summary of the type of documents. Different files will have different contents which will influence the performance of the compression result. Similar to the natural images having different conditions such as illumination, scenes, objects and so on, screen contents have various types of documents and softwares. The x-axis is the type of documents and the y-axis is the number of each type. 
    }
    \label{fig:docs_type}
\end{figure}
After the subjective analysis, we want to show some analysis from the objective part to support that our proposed dataset covers a wide data distribution of the scene data. As we all know, one of the key assumptions of the supervised method is that the training dataset can represent the real distribution. So we want to use some objective analysis to support that the proposed dataset is useful in future's research.

\begin{figure}
    \centering
    \includegraphics[height=0.3\textwidth]{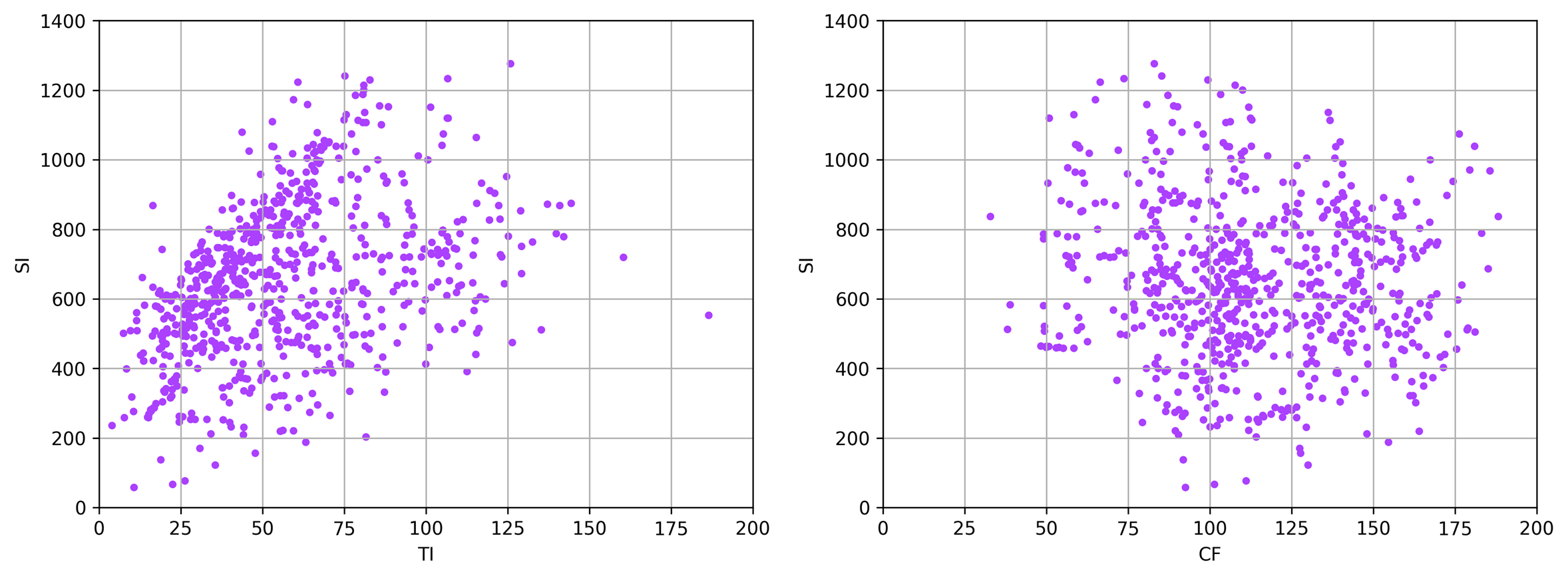}
    \caption{The scatter plots of the video descriptors in the LSCD. The left one shows the the TI(Temporal information) and SI(Spatial information) and the right one shows the CF(Colorfulness) and SI(Spatial information).}
    \label{fig:siti}
\end{figure}

Firstly, we want to give you an analysis of the document types. \cref{fig:docs_type} shows the distribution of the document types of the dataset. As we can know from the dataset, we try to collect various document types in the daily use of the computer. And we use these to compose different scenarios to reflect the daily usage of the computer.
Secondly, we show the SI-TI and  SI-CF in \cref{fig:siti}. Spatial information(SI) and Temporal information(TI)\cite{rec2008p} are usually used for reflecting the distribution of the video data. 
Colorfulness(CF)\cite{winkler2012analysis} represents a perceptual indicator of the variety and intensity of colors. And from this figure, we can know that the proposed dataset has a good distribution under these descriptors.
\section{Experiments of Building the Benchmark}
For making the benchmark of LSCD, we separate the current methods into two parts, the traditional and the deep learning-based methods. For the traditional methods, we select the common codecs such as H.264\cite{wiegand2003overview}, H.265\cite{sullivan2012overview} and etc. For the deep learning methods, we choose some representative methods and directly test them on the \textit{Testing Set} and the \textit{Additional Testing Set}. For evaluation metrics, we use PSNR and MS-SSIM to evaluate reconstruction quality, and bits per pixel (bpp) for compression ratio for both traditional codecs and learning-based methods.
\subsection{Setting up}
For traditional codec, we use the software FFmpeg\cite{ffmpeg} to do the experiments and we use the build\cite{ffmpeg_build} which contains all the libraries we need. And all the traditional codec tests are running on the CPU-server. 
For deep-learning based methods, we use the GPU-server with V100 to run the experiments. And the detail information is shown in the supplemental material.
\subsection{Traditional Codec}
\begin{figure}
    \centering
    \includegraphics[width=0.8\linewidth]{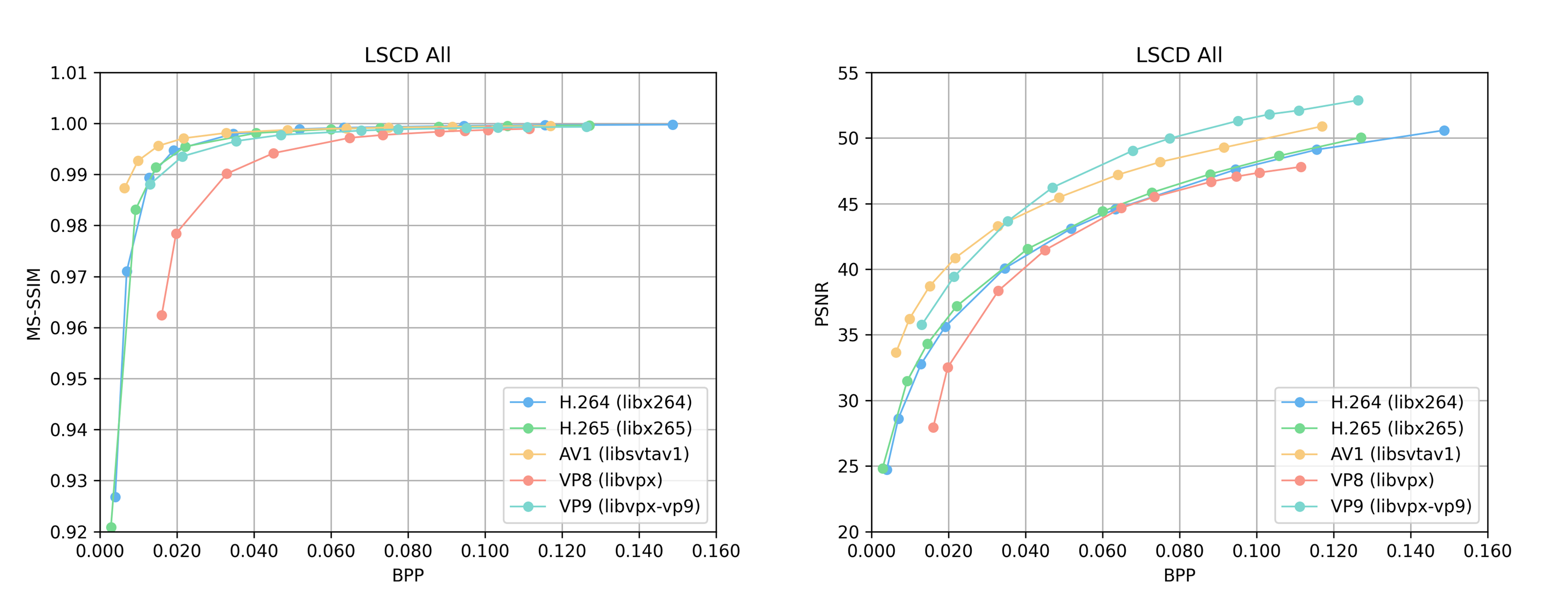}
    \caption{The performance of traditional codecs on the proposed dataset(Better in color). The left one is the performance of the MS-SSIM and the right one is the performance of PSNR. And for each codec, the implementations of them are in the parentheses. }
    \label{fig:codec_preformance}
\end{figure}

The results of LSCD are shown in \cref{fig:codec_preformance}. The traditional codecs we choose are H.264\cite{wiegand2003overview}, H.265\cite{sullivan2012overview}, VP8\cite{bankoski2011technical}, VP9\cite{mukherjee2015technical} and AV1\cite{han2021technical}. And for each codec, we select one implementation to do the experiments instead of using the reference software. However, we do not use the H.266\cite{bross2021overview}, which is the latest codec. The reason is that we did not find a good third-party implementation of that, and the reference software of H.266 is so slow that we can not do experiments on our proposed large-scale dataset. 
\subsection{Learning-based Methods}
Not only for the traditional codec, but we also test some representative deep learning methods on the test part of the proposed dataset. The representative methods are individually for one category of methods. DVC\cite{lu2019dvc} and LHBDC\cite{yilmaz2021end} are for the type which replaces each module in the traditional method. DCVC-DC\cite{li2023neural} is for the method that leverages the RNN-style structure to increase the both temporal and spatial context diversity so as to improve the compression performance. And NeRV\cite{chen2021nerv} is the implicit neural representation for videos which aims to overfit each data sample and then turn the compression of the video problem into the problem of compressing the models. Although the learning-based methods could achieve a comparable performance in quality and compression, they need much more time than the traditional codec even if using the GPUs. And the detail information and analysis are provided in the supplemental materials.
The \cref{fig:dl_performance} shows these methods' performance on the test part of the proposed dataset because we leave out the training set for future research. 
\begin{figure}
    \centering
    \includegraphics[height=0.6\textwidth]{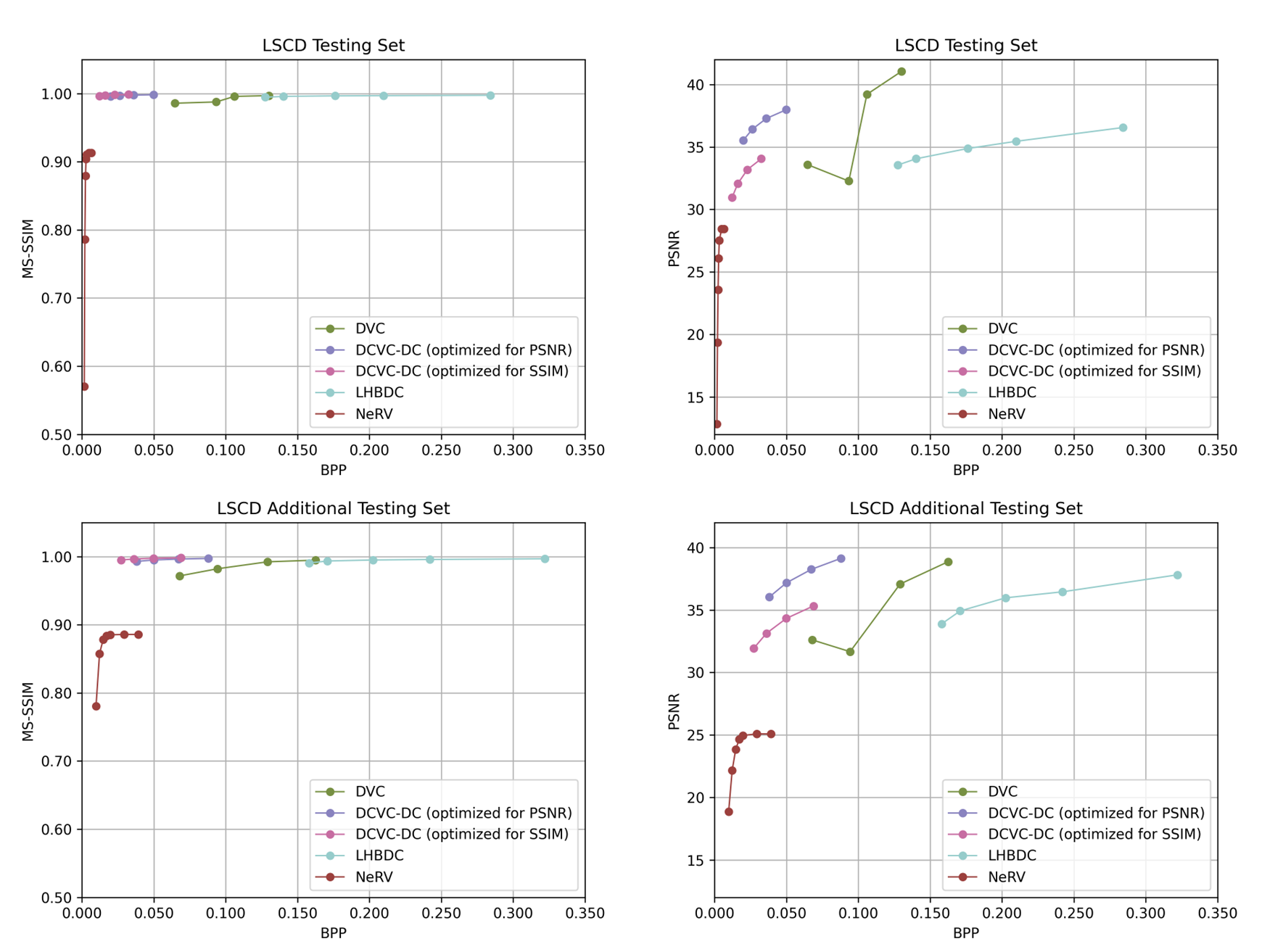}
    \caption{The performance of deep learning based methods on the proposed dataset(Better in color).}
    \label{fig:dl_performance}
\end{figure}
\section{Technical Limitation}
\label{sec:tech_limit}
Hence we now outline some potential limitations of the proposed dataset. Meanwhile, these limitations may be the future work to continue building the dataset.

\textbf{Limitation 1.} Not including all actual usage of scenes. There are so many scenes that users could use the screen in daily life, and it is also impossible that we could collect all the scenes. For example, we do not include the gaming scenes. The reasons are from two aspects. One is that we think that gaming is a so large type of scenarios that could be an individual topic to research, because it will need less delay and higher resolution. The other one is that if we record the gaming scene, the requirement of devices is higher than usual. 

\textbf{Limitation 2.} Not including the fine-tuned methods to build the benchmark. Following the normal form of the computer vision, we need fine tune the previous models on the training set to get a better performance on the testing set. Nevertheless, we think the benchmark should reflect the actual results of the previous methods instead of chasing the high performance. Meanwhile, this also shows the generalization of the previous methods. 

\textbf{Limitation 3.} Not including the high resolution and FPS videos. Although we make a hypothesis that most common setting is 1080p and 25 FPS, we can not deny that there are high resolutions and FPS. And at present, LSCD does not include these data, so if researchers use the model trained by this dataset in the high resolution and FPS, they may need do some pre-process or post-process to achieve a better performance.
\section{Conclusion}
With the rapid development of mobile devices, screen content videos have become emerging visual data, and we need to compress them more frequently than before. However, there is a blank for a large-scale dataset for both traditional codecs and learning-based methods. So, in this paper, we propose a new large-scale screen content dataset for video compression. Meanwhile, we describe the process of building the dataset, analyze the proposed dataset from both subjective and objective aspects, and do the experiments of both traditional and learning-based methods to provide a benchmark for future research.  
In the future, we will enlarge this dataset for more areas of compression, such as visual quality assessment and so on. And we will continue our work to contrapose the current technical limitations.  

\clearpage
\bibliographystyle{unsrt}
\bibliography{refs}
\clearpage

\appendix
\title{Appendix}
\section{Motivation}
This dataset is proposed to provide a huge number of lossless screen content video data mainly for video compression and promote the development of technologies in both traditional and learning-based methods.

\section{Uses}
The LSCD dataset aims to advance the development of screen content video compression and makes it more clear about the performance of traditional and learning-based methods nowadays, i.e., the proposed benchmark. Besides being applied in video compression, this dataset can also be used for image compression, image restoration, and other tasks related to images. But it might be more difficult for other computer vision tasks, such as image segmentation, video tracking, and object detection.

\section{Composition}
The LSCD dataset consists of 714 distinct sequences, which contain 172M frames in total, covering the vast majority of usage scenarios, including browsing webpages, dynamic desktops, viewing geological maps, etc. Each frame is stored in BMP format, but because of the huge demand of storage capacity, about 10 TB, we only release the lossless dataset with PNG format, roundly 1 TB. Feel free to contact us to get BMP format samples. Detailed content of each level part in LSCD dataset is shown below.
\begin{itemize}
    \item \textbf{Plain}: this part contains 30 sequences, including viewing 5 word documents, 3 slides, 2 excel documents, 2 PDFs, 2 \TeX\ documents, 2 text documents, watching 6 different types of videos, browsing 4 various webpages, and 4 dynamic desktops.
    \item \textbf{Medial}: besides the normal usage in plain part, the medial part includes very common usage scenarios, 281 in total, such as editing visio documents, utilizing Adobe Photoshop to adjust images and Adobe Effects to edit videos, playing music via Spotify and other software, viewing abundant geological maps, and the most often used, online conferences.
    \item \textbf{Complex}: this part, including 293 sequences, aims to simulate multi-task conditions in daily life. When you processing multiple tasks, you usually have multiple windows on your desktop. We try our best to make 293 such conditions, such as browsing webpages and communicating with your colleagues, searching information on webpage and editing slides, viewing codes of different projects, and so on.
    \item \textbf{Supp}: this part contains 110 sequences, designed for fast motion and/or time-sync comments, namely bullet chatting. We make 14 sequences of dragging windows quickly, 24 sequences of scrolling windows fast, 30 sequences with time-sync comments on Bilibili, 30 sequences without time-sync comments on YouTube, and 10 sequences of scene changing. Notably, the sequences on Bilibili and YouTube are split into fullscreen, theatre mode, and playing with small window on the webpage.
\end{itemize}
\section{More information related to the LSCD}
\subsection{Visualization}
\begin{figure}
    \includegraphics[width=\linewidth]{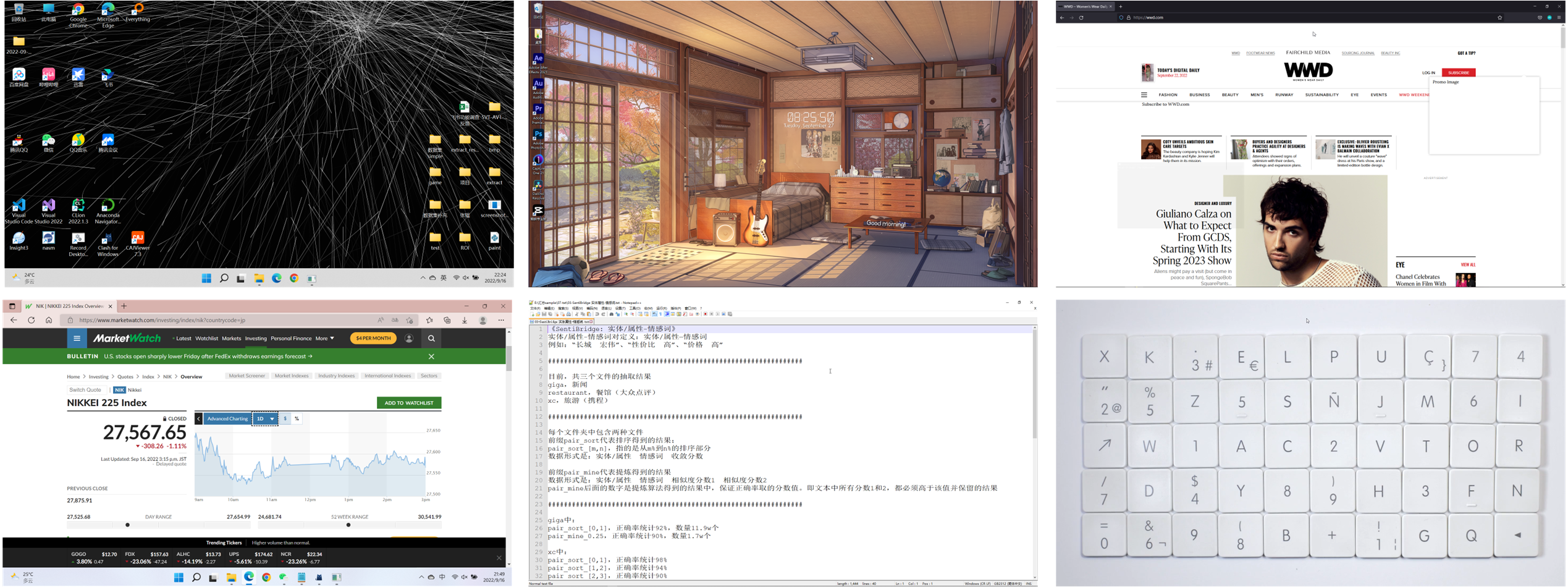}
    \caption{More visualizations of the plain part.}
    \label{fig:plain}
\end{figure}

\begin{figure}
    \includegraphics[width=\linewidth]{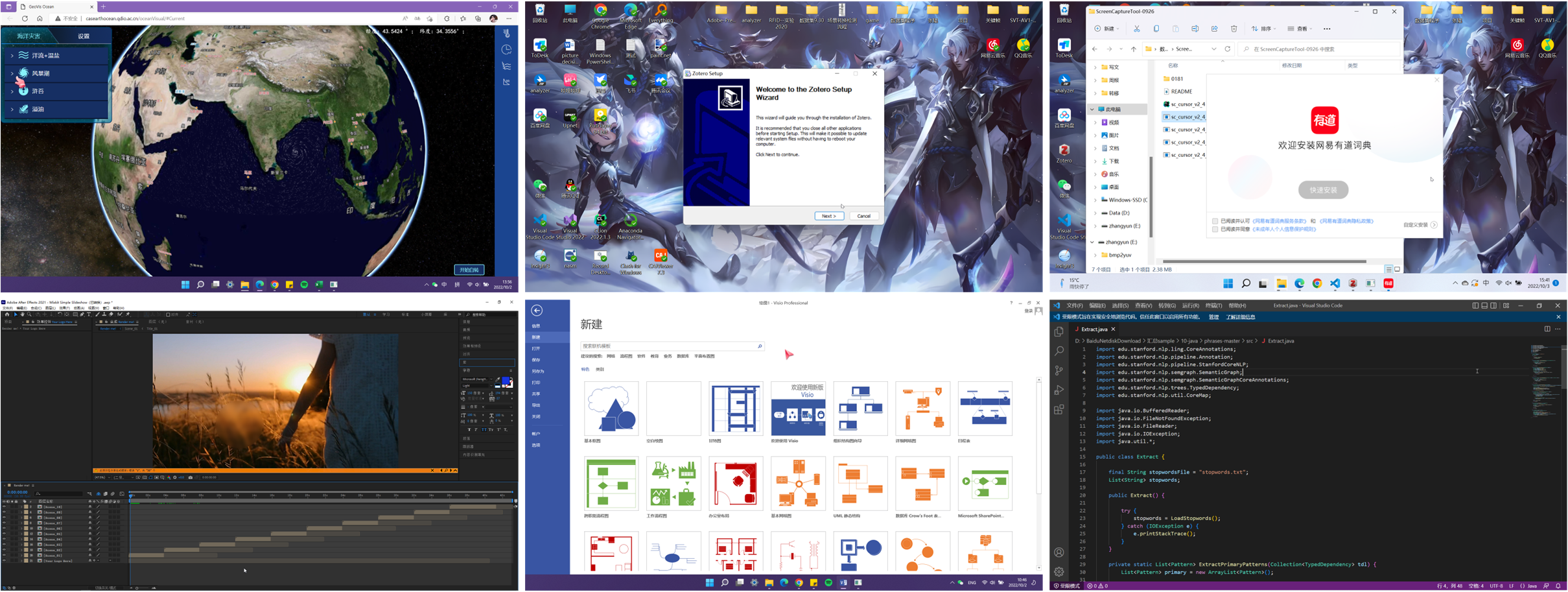}
    \caption{More visualizations of the medial part.}
    \label{fig:medial}
\end{figure}

\begin{figure}
    \includegraphics[width=\linewidth]{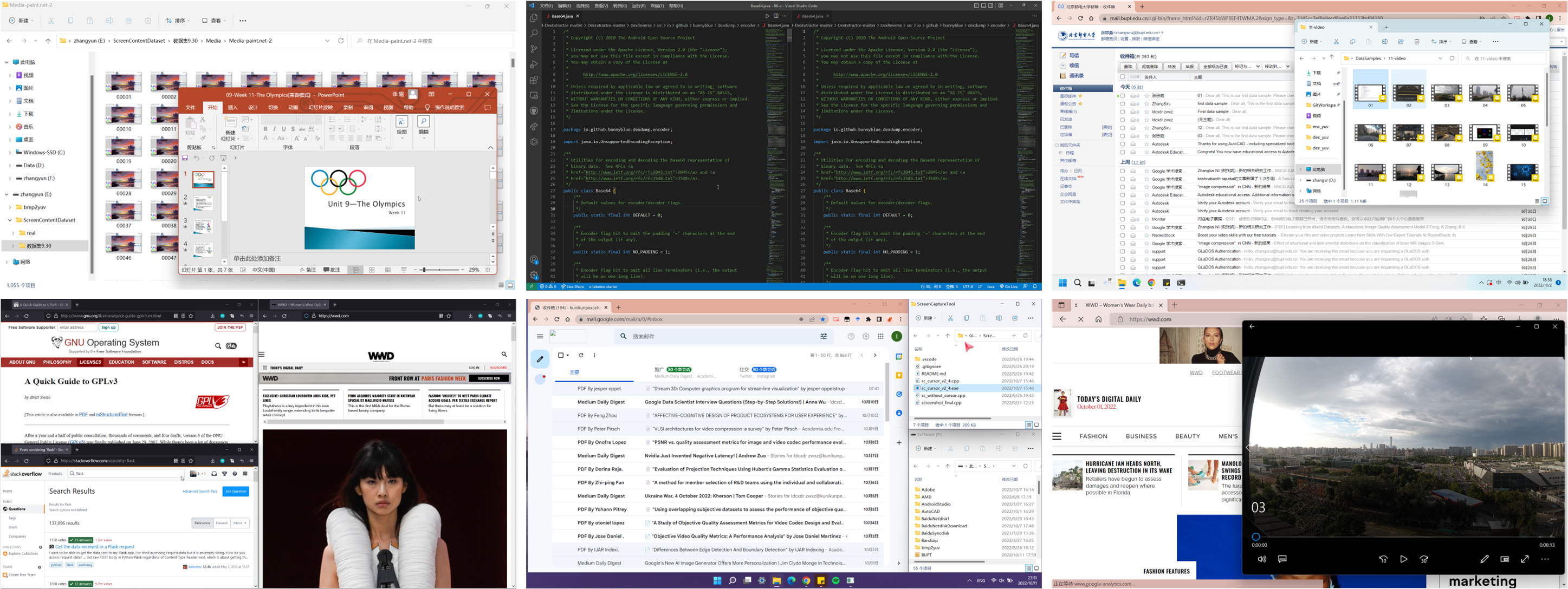}
    \caption{More visualizations of the complex part.}
    \label{fig:complex}
\end{figure}

\begin{figure}
    \includegraphics[width=\linewidth]{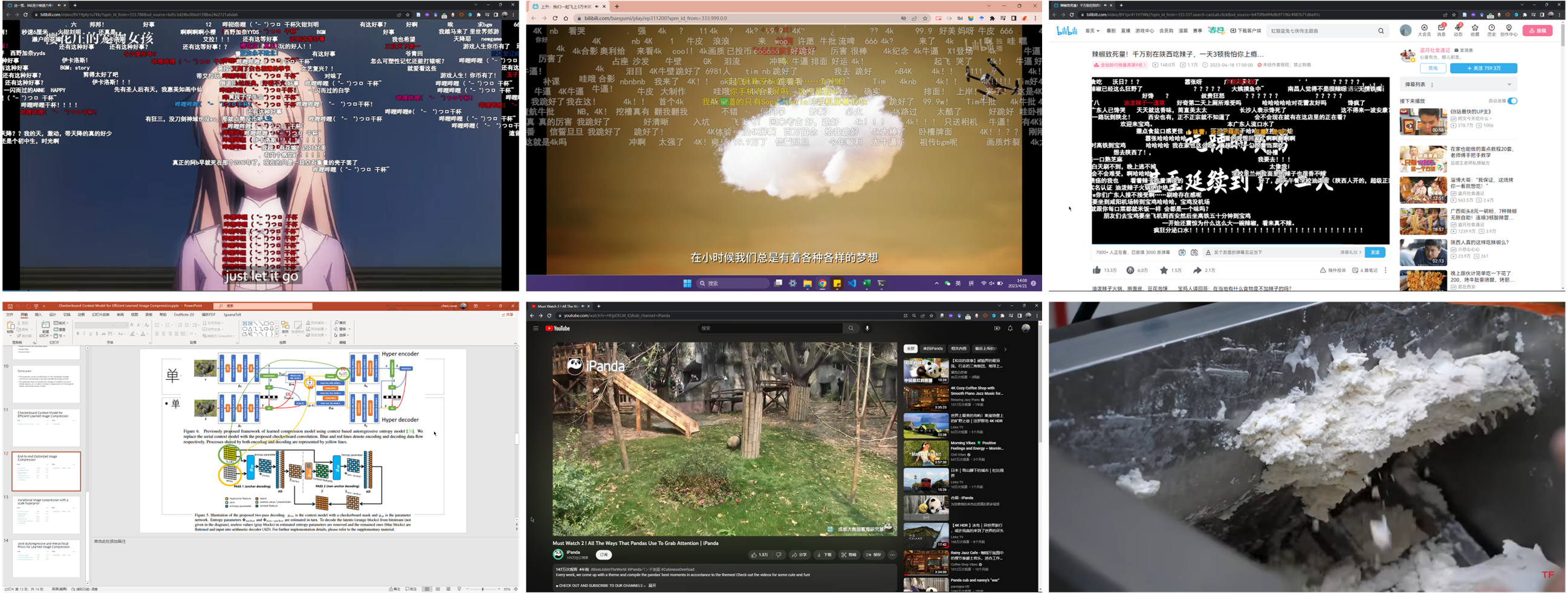}
    \caption{More visualizations of the supp part.}
    \label{fig:supp}
\end{figure}

In the main body part, we have shown part of the thumbnail of the proposed dataset which is just a small part of the whole dataset. And in this part we will show you more samples in our dataset, however, this still a part of the whole dataset. If you want to explore the whole LSCD, we suggest you get the whole dataset. Meanwhile, we show the demo of each part. \cref{fig:plain}, \cref{fig:medial}, \cref{fig:complex} and \cref{fig:supp} show the more visualizations for plain, medial, complex and supp part respectively.

\subsection{Analyzing}
\begin{figure}
    \includegraphics[width=\linewidth]{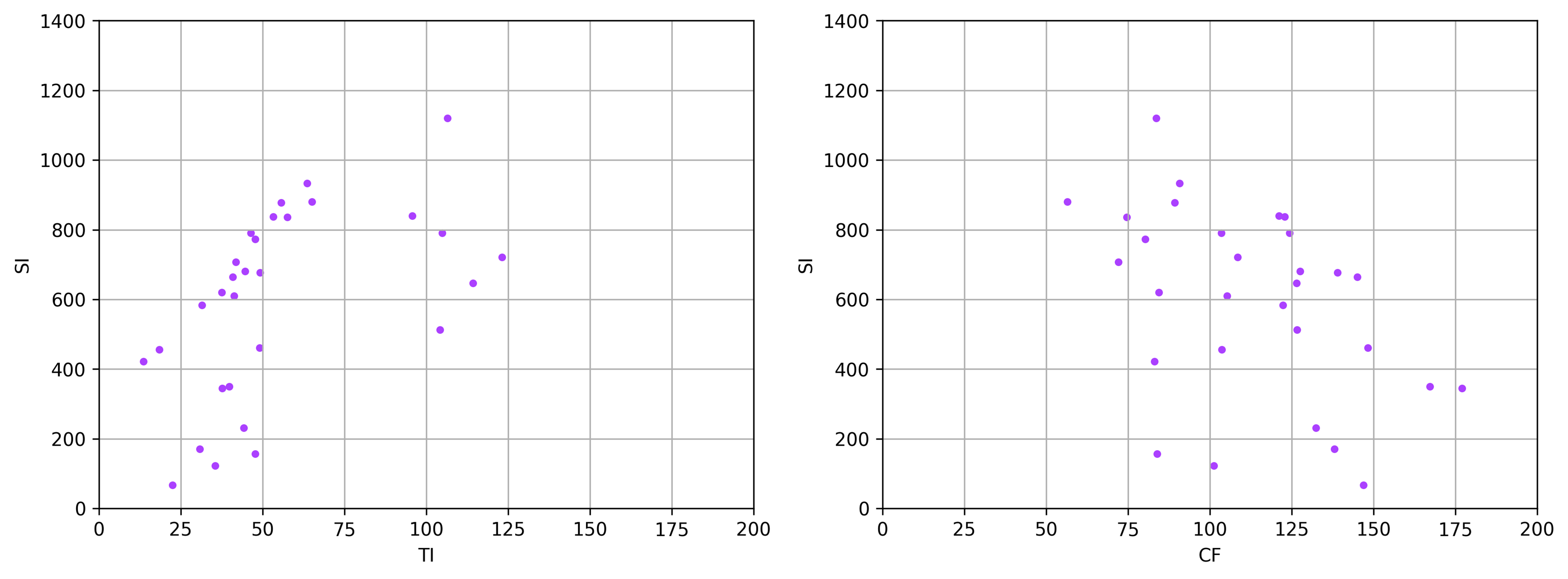}
    \caption{The video descriptors in plain part.}
    \label{fig:plain_siticf}
\end{figure}

\begin{figure}
    \includegraphics[width=\linewidth]{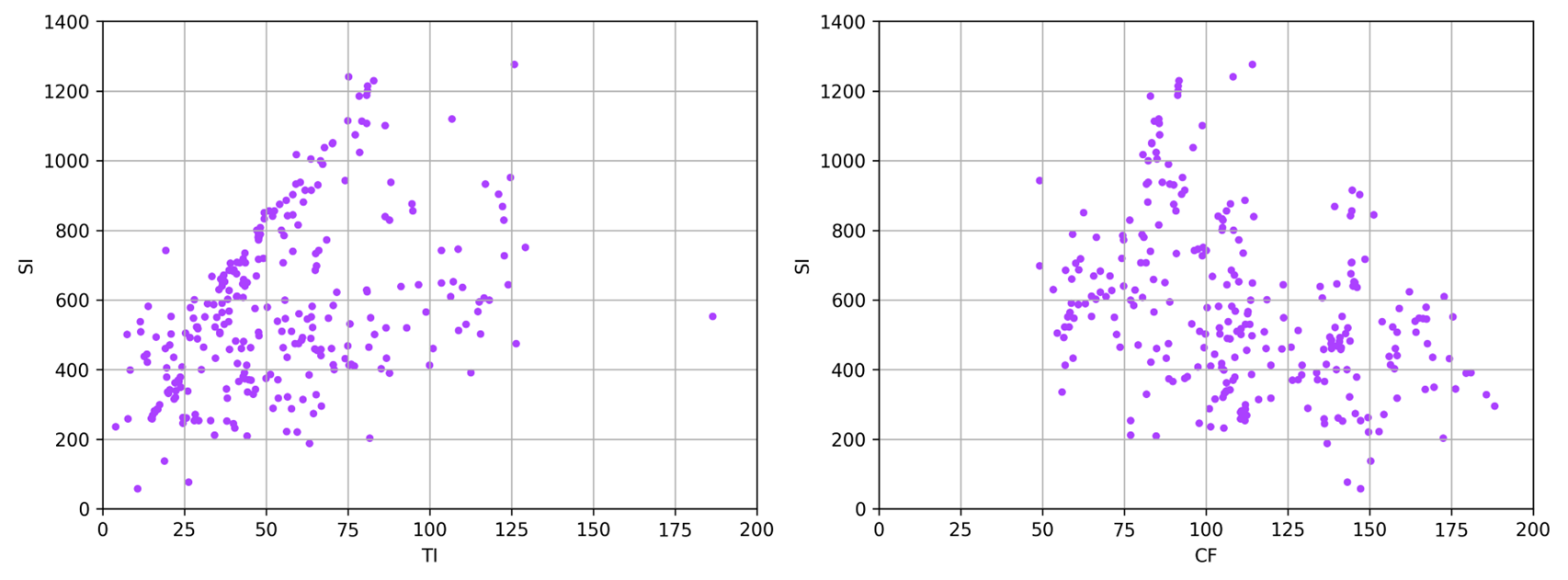}
    \caption{The video descriptors in medial part.}
    \label{fig:medial_siticf}
\end{figure}

\begin{figure}
    \includegraphics[width=\linewidth]{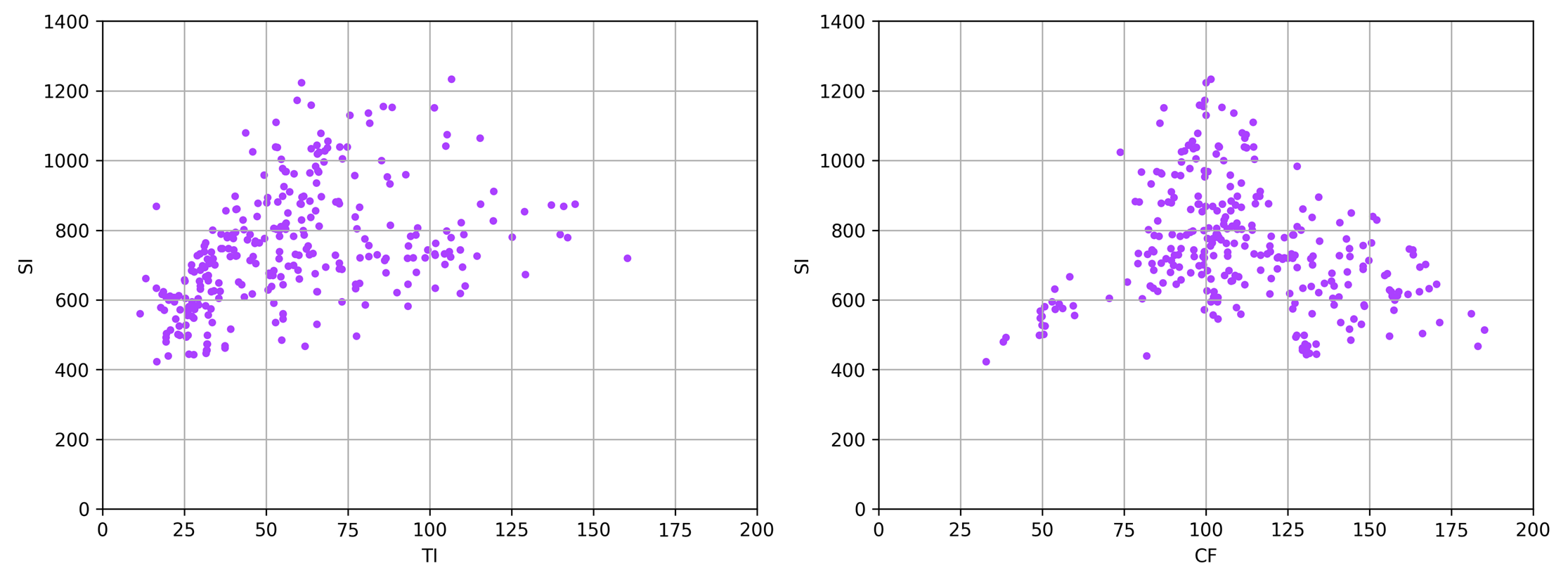}
    \caption{The video descriptors in complex part.}
    \label{fig:complex_siticf}
\end{figure}

\begin{figure}
    \includegraphics[width=\linewidth]{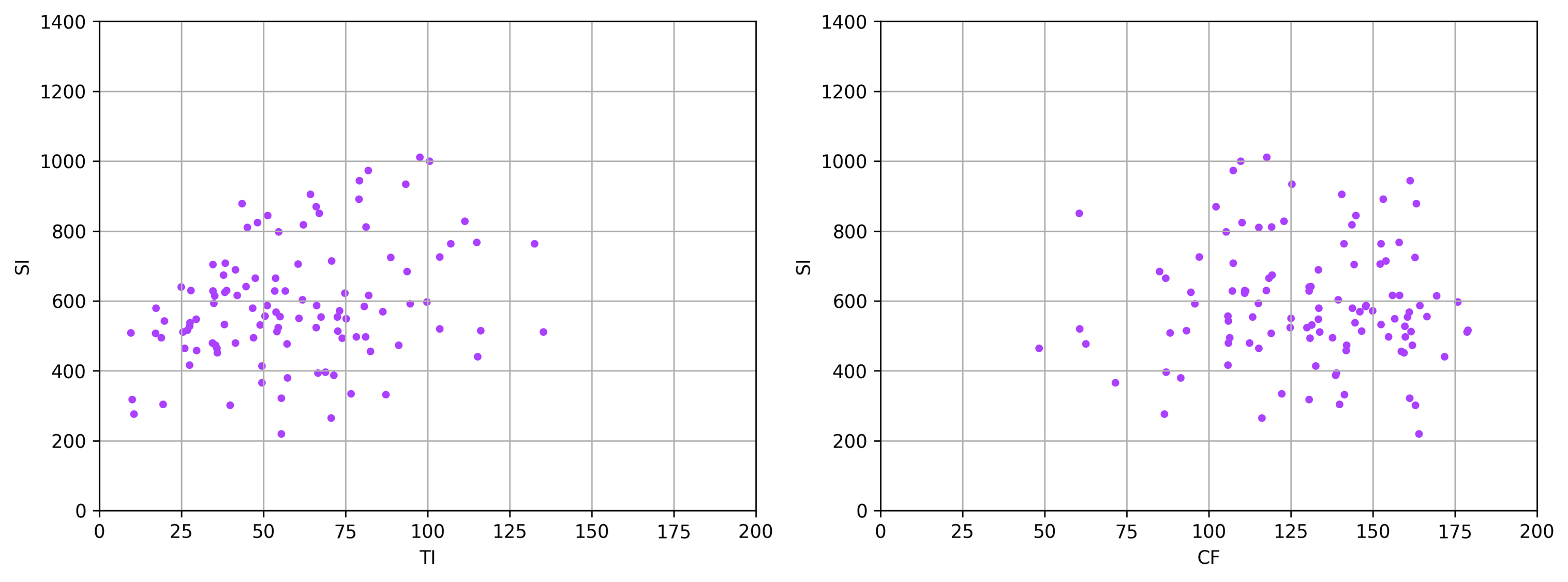}
    \caption{The video descriptors in supp part.}
    \label{fig:supp_siticf}
\end{figure}

In this part, we will show more figures related to the SI(Spatial Information), TI(Temporal Information) and CF(Colorfulness) of each part in the proposed dataset. \cref{fig:plain_siticf}, \cref{fig:medial_siticf}, \cref{fig:complex_siticf} and \cref{fig:supp_siticf} show these stuff for each part respectively. From these figures, we know that each part of the dataset has a good distribution of both SI-TI and SI-CF. And, as a result of the different number of videos, we have a different density of points of each part.

\section{The details of the experiments}
Except the figures in the main body, we also provide you the experiments of each part of the proposed LSCD. In this way, we want to help researchers to explore the algorithms better on each individual part.
\subsection{Traditional codecs}
\begin{figure}
    \includegraphics[width=\linewidth]{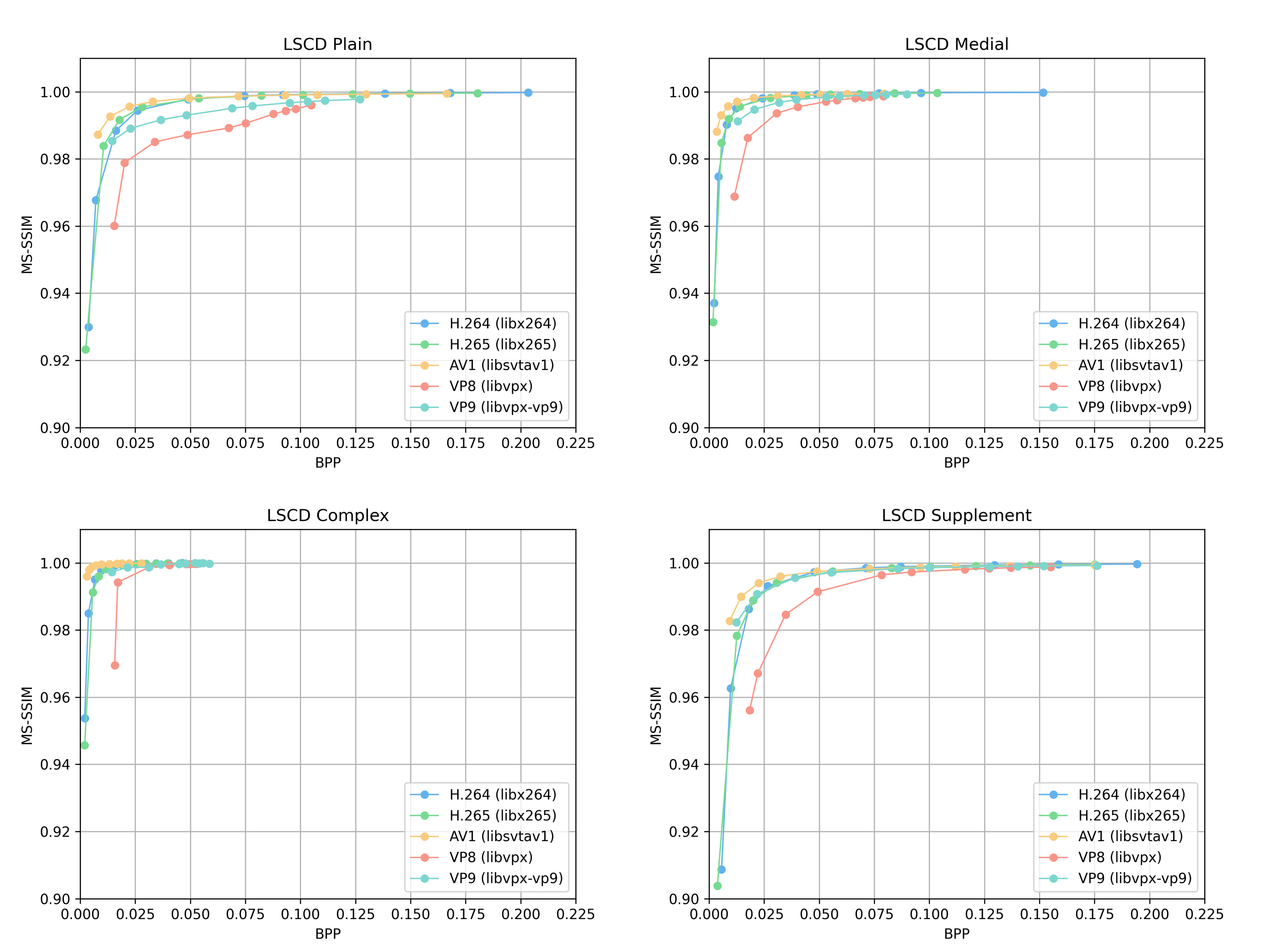}
    \caption{The traditional codecs' MS-SSIM performance in each LSCD level part.}
    \label{fig:trad_ssim_eachpart}
\end{figure}

\begin{figure}
    \includegraphics[width=\linewidth]{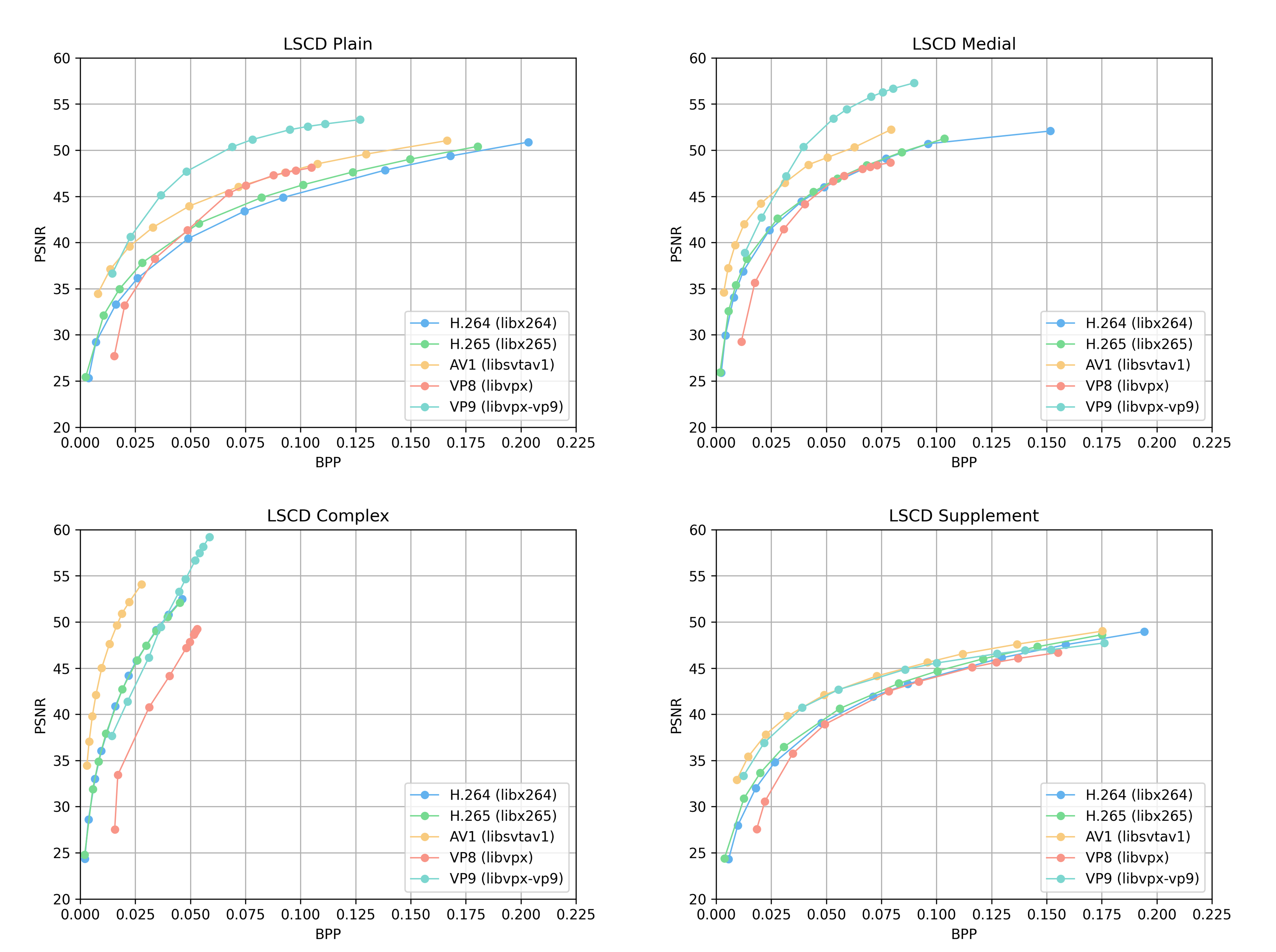}
    \caption{The traditional codecs' PSNR performance in each LSCD level part.}
    \label{fig:trad_psnr_eachpart}
\end{figure}

This part shows the detail experiments result of traditional codecs in each LSCD part. \cref{fig:trad_ssim_eachpart} shows these traditional codecs' performance under the measure of the MS-SSIM metric, and \cref{fig:trad_psnr_eachpart} shows that of the PSNR metric. From these figures, we find that the complex part has the better performance than the plain and the medial part, which is a surprising result. 

\subsection{Learning-based Methods}
\begin{figure}
    \includegraphics[width=\linewidth]{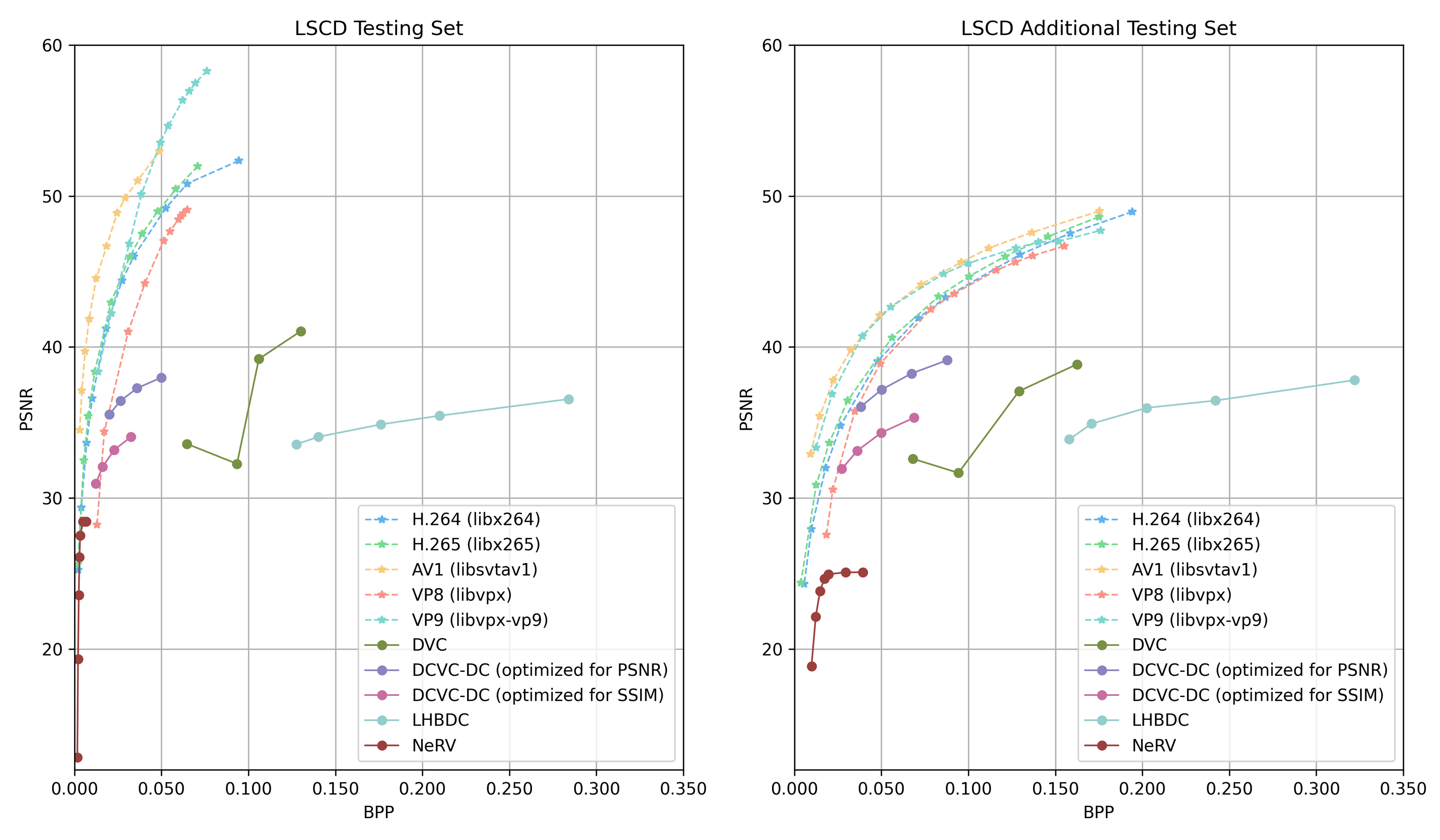}
    \caption{The PSNR of traditional and learning-based methods.}
    \label{fig:trad_dl_psnr}
\end{figure}

\begin{figure}
    \includegraphics[width=\linewidth]{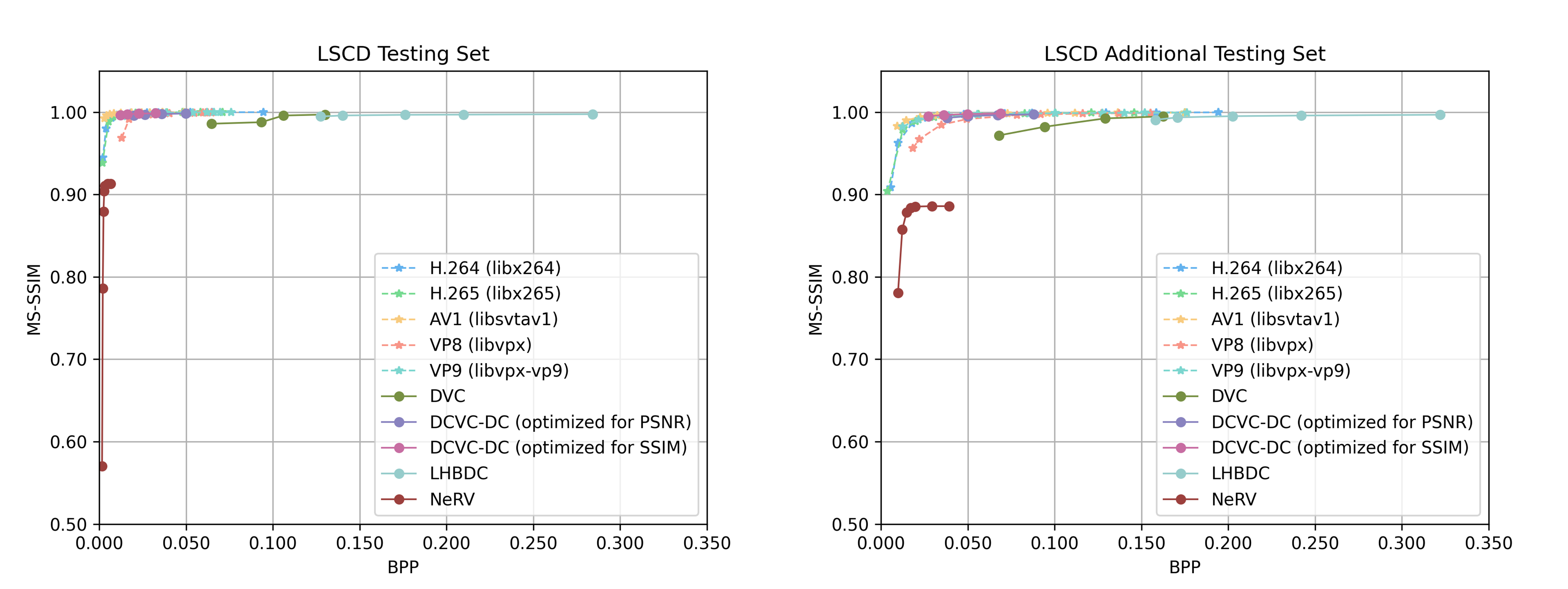}
    \caption{The MS-SSIM of traditional and learning-based methods.}
    \label{fig:trad_dl_ssim}
\end{figure}

This part will introduce the experiments results of the comparison between the traditional codec and the learning-based methods on the same part, the \textit{Testing Set} and \textit{Additional Testing Set}. As we have mentioned in the main body, the \textit{Testing Set} consists of part of plain, medial and complex videos and \textit{Additional Testing Set} is the supplemental videos. \cref{fig:trad_dl_psnr} and \cref{fig:trad_dl_ssim} show this comparison. 

\begin{figure}
    \includegraphics[width=\linewidth]{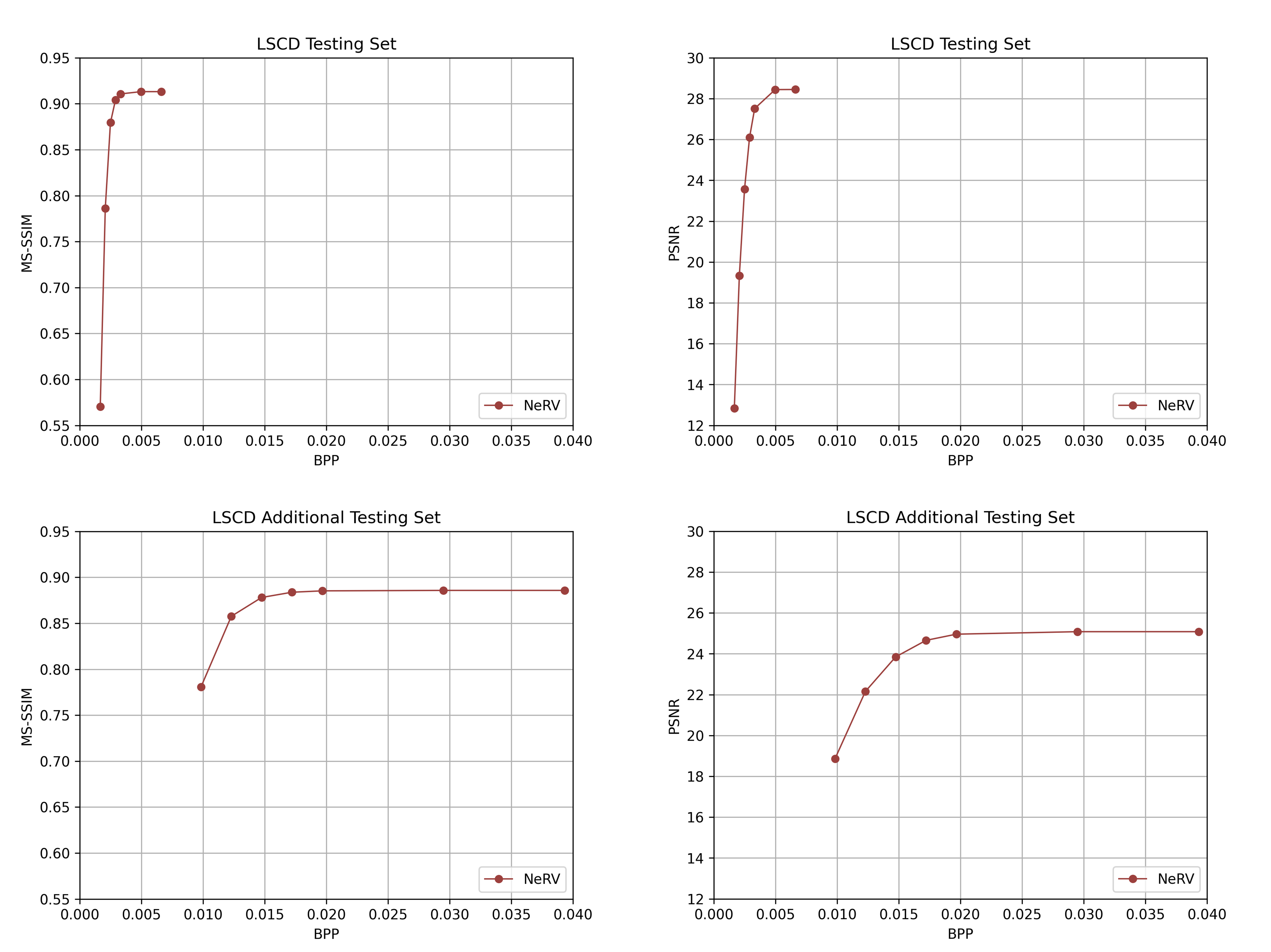}
    \caption{The performance of the NeRV in \textit{Testing Set} and \textit{Additional Testing Set}.}
    \label{fig:nerv}
\end{figure}

Moreover, we individually provide the NeRV's performance in \cref{fig:nerv}. The reason why we provide this figure is that we find the NeRV's performance is so compact on the low-bpp area and we want to make it clear for the readers. 

\subsection{The time of deep learning methods}
\begin{table}[]
    \caption{The time summary of the learning-based methods. \textit{Time} entry means the total time of each model encoding and decoding one frame. For the NeRV method, we show it's time of inference. }
    \centering
    \begin{tabular}{c|ccccc}
    \hline
            & DCVC-DC$_{psnr}$\cite{li2023neural} & DCVC-DC$_{ssim}$\cite{li2023neural} & LHBDC\cite{yilmaz2021end} & NeRV\cite{chen2021nerv} & DVC\cite{lu2019dvc}  \\ \hline
    Time(s) & 0.90         & 0.85         & 6.19  & 0.03           & 2.90 \\ \hline
    \end{tabular}
    \label{tab:dl_time}
\end{table}

\begin{table}[]
    \caption{The time of NeRV's different phases.The training process includes the training and pruning on the individual data sample and the inference means the model reconstructs the frames.The time means the average time of all of the frames. }
    \centering 
    \begin{tabular}{c|cc}
    \hline
            & Training Process & Inference Process \\ \hline
    Time(s) & 0.31             & 0.03              \\ \hline
    \end{tabular}
    \label{tab:nerv_train}
\end{table}

\cref{tab:dl_time} shows the inference time of each learning-based methods for one frame. From this table, we can know that the time of the learning-based methods are longer than we imagine even if they use the GPUs to accelerate the process. And \cref{tab:trad_time} shows the time of traditional codecs. From the comparison between these two tables, we can get a more clear about the difference. So we think how to reduce the time is much more important than improving the performance of the learning-based methods. Specifically, NeRV is different from the other learning-based methods and it needs training on each video. And \cref{tab:dl_time} just shows the time of inference which does not include the training process, while \cref{tab:nerv_train} shows the time of its training process.
\begin{table}[]
    \centering
    \caption{The time of the traditional codecs.This table shows the encoding time, decoding time and the total time of each traditional codec. As we have mentioned, we use \textit{libx264} for H.264, \textit{libx265} for H.265, \textit{libsvtav1} for AV1, \textit{libvpx} for VP8 and \textit{libvpx-vp9} for VP9.}
    \begin{tabular}{c|ccccc}
    \hline
                & H.264 & H.265 & AV1 & VP8 & VP9 \\ \hline
    Encode Time(s) & 0.0404  & 0.0921  & 0.0870    & 0.4442 & 0.4796     \\
    Decode Time(s) & 0.0242  & 0.0305  & 0.0305    & 0.0292 & 0.0295     \\
    Total Time(s)  & 0.0647  & 0.1226  & 0.1175    & 0.4734 & 0.5092     \\ \hline
    \end{tabular}
    \label{tab:trad_time}
\end{table}
\section{The settings of the traditional codecs}
This part will introduce some basic settings when we do the experiments of the traditional codecs. We use the FFmpeg\cite{ffmpeg} to do the experiments, and in order to make sure we do not have any errors during the process of building the FFmpeg, we use the built version in \cite{ffmpeg_build}.
\subsection{H.264}
We use the \textit{libx264} built in FFmpeg to do experiments of the H.264. The explicit commands is as following:
\begin{verbatim}
ffmpeg -y -f rawvideo -pix_fmt yuv420p -s:v 1920x1080 -framerate 25 -i in.yuv 
        -c:v libx264 -crf CRF -preset medium 
        -x264-params bframes=0 out.h264
\end{verbatim}
The \textbf{CRF} is a variable whose values are: 12, 14, 16, 20, 22, 26, 32, 36, 42, 48.
\subsection{H.265}
We use the \textit{libx265} built in FFmpeg to do experiments of the H.265. The explicit commands is as following:
\begin{verbatim}
ffmpeg -y -f rawvideo -pix_fmt yuv420p -s:v 1920x1080 -framerate 25 -i in.yuv 
        -c:v libx265 -crf CRF -preset medium 
        -x265-params bframes=0 out.hevc
\end{verbatim}
The \textbf{CRF} is a variable whose values are: 14, 16, 18, 20, 22, 26, 32, 36, 40, 50.
\subsection{VP8}
We use the \textit{libvpx} built in FFmpeg to do experiments of the VP8. The explicit commands is as following:
\begin{verbatim}
ffmpeg -y -f rawvideo -pix_fmt yuv420p -s:v 1920x1080 -framerate 25/1 -i in.yuv 
        -c:v libvpx -tune psnr 
        -b:v BITRATE -minrate BITRATE -maxrate BITRATE out.ivf
\end{verbatim}
The \textbf{BITRATE} is a variable whose values are: 500K, 1M, 2M, 3M, 5M, 6M, 8M, 9M, 10M, 12M.
\subsection{VP9}
We use the \textit{libvpx-vp9} built in FFmpeg to do experiments of the VP9. The explicit commands is as following:
\begin{verbatim}
ffmpeg -y -f rawvideo -pix_fmt yuv420p -s:v 1920x1080 -framerate 25/1 -i in.yuv 
        -c:v libvpx-vp9 -deadline realtime -tune psnr 
        -b:v BITRATE -minrate BITRATE -maxrate BITRATE out.ivf
\end{verbatim}
The \textbf{BITRATE} is a variable whose values are: 500K, 1M, 2M, 3M, 5M, 6M, 8M, 9M, 10M, 12M.
\subsection{AV1}
We use the \textit{libsvt-av1} built in FFmpeg to do experiments of the AV1. The explicit commands is as following:

\begin{verbatim}
ffmpeg -y -f rawvideo -pix_fmt yuv420p -s:v 1920x1080 -i in.yuv 
        -c:v libsvtav1 -framerate 25/1 -crf CRF 
        -svtav1-params lookahead=0:hierarchical-levels=3:preset=11 out.ivf
\end{verbatim}

The \textbf{CRF} is a variable whose values are: 10, 14, 18, 22, 28, 35, 42, 48, 55, 60.
\section{The settings of the deep learning based methods}
As we have mentioned in the main body, we do the experiments of DVC\cite{lu2019dvc}, LHBDC\cite{yilmaz2021end}, DCVC-DC\cite{li2023neural} and NeRV\cite{chen2021nerv} on the \textit{Testing Set} and the \textit{Additional Testing Set}. And this part will introduce the settings when we doing the experiments of deep learning methods.
\subsection{Hardware and Software Settings}
The experiments are executed on the GPU Servers and each experiment is running on the single GPU which is V100 with 32GB memory. For the software, each learning-based method may have some difference, so you can check the requirements of them and follow the instructions to install the required softwares. 
\subsection{Commands}
This section introduces the commands to run the learning-based methods. 

\subsubsection{DVC}
For DVC, we run the inference process with $batchsize=4$, $workers=8$ on 4 pretrained model with $lambda=256, 512, 1024, 2048$, respectively. The example command of \textit{Additional Testing Set} is as following:
\begin{verbatim}
python -u savecode/main.py --model_version dvc --batch_size 4 --num_workers 8 
                --LSCD_supp_dataset_path image_1920x1024/supp/ 
                --video_list_LSCD_supp image_1920x1024/video_list_supp.txt 
                --test_LSCD_supp --train_lambda 256 
                --pretrain snapshot/pretrain_256.model
\end{verbatim}

\subsubsection{LHBDC}
For LHBDC, we run the inference process with $batchsize=1$ and $workers=4$ with $lambda=228, 436, 845, 1626, 3141$, respectively. The example command of \textit{Additional Testing Set} is as following:
\begin{verbatim}
python testing.py --test_gop_size 8 --i_interval 8 --i_qual 7 --workers 4
            --lmbda_list 228 436 845 1626 3141 
            --test_path LSCD/supp 
            --test_lscd_list dataset_subset/supp.csv  
\end{verbatim}

\subsubsection{DCVC-DC}
For DCVC-DC, we run the inference process with $rate\_num=4$, i.e., calculating 4 rate points once, and $worker=1$. DCVC-DC is optimized for both PSNR and SSIM. The example command for PSNR model is as following:
\begin{verbatim}
python test_video.py --rate_num 4 --test_config ./LSCD_plain.json --yuv420 0
            --i_frame_model_path ./checkpoints/cvpr2023_image_psnr.pth.tar 
            --p_frame_model_path ./checkpoints/cvpr2023_video_psnr.pth.tar 
            --cuda 1 --worker 1 --write_stream 0 
            --output_path output_plain_psnr.json 
            --force_intra_period 32 --calc_ssim 1 --verbose 0
\end{verbatim}
The example command for SSIM model is as following:
\begin{verbatim}
python test_video.py --rate_num 4 --test_config ./LSCD_plain.json --yuv420 0
            --i_frame_model_path ./checkpoints/cvpr2023_image_ssim.pth.tar 
            --p_frame_model_path ./checkpoints/cvpr2023_video_ssim.pth.tar 
            --cuda 1 --worker 1 --write_stream 0 
            --output_path output_plain_ssim.json 
            --force_intra_period 32 --calc_ssim 1 --verbose 0
\end{verbatim}

\subsubsection{NeRV}
For NeRV, we need to pre-train, prune, and evaluate the pruned model for each sequence.
For pretrain process, we set $batchsize=6$, $strides=5, 3, 2, 2, 2$ for 1920x1080 resolution, $embed=1.25\_80$ as NeRV recommended, and $epoch=300$ as constraint of training time. The example command of Testing Set is as followings:
\begin{verbatim}
python train_nerv.py -e 300 --lower-width 96 --num-blocks 1 --frame_gap 1 
            --embed 1.25_80 --stem_dim_num 512_1  --reduction 2  
            --fc_hw_dim 9_16_26 --expansion 1 --single_res 
            --loss Fusion6 --warmup 0.2 --lr_type cosine 
            --strides 5 3 2 2 2 --conv_type conv -b 6 --lr 0.0005 --norm none 
            --act swish --dataset LSCD/complex --lscd_level complex 
            --test_lscd_list dataset_subset/complex.csv
\end{verbatim}
For inference process, the parameters are the same as that of pretrain process, except that $prune\_ratio=0.4$ as a good trade-off between accuracy and efficiency, $warmup=0.$, and $epoch=50$. The example command is as following:
\begin{verbatim}
python train_nerv.py -e 50 --lower-width 96 --num-blocks 1 --frame_gap 1 
            --embed 1.25_80 --stem_dim_num 512_1 
            --reduction 2 --fc_hw_dim 9_16_26 --expansion 1 
            --single_res --loss Fusion6 --warmup 0. --lr_type cosine 
            --strides 5 3 2 2 2 --conv_type conv -b 6 --lr 0.0005 --norm none 
            --act swish --prune_ratio 0.4 --not_resume_epoch 
            --dataset LSCD/complex --lscd_level complex 
            --test_lscd_list dataset_subset/complex.csv --autoresume
\end{verbatim}
For inference process, the parameters are the same as that of prune process, except that $batchsize=1$, $quant\_axis=0$, and adding $quant\_bit\_list=4, 5, 6, 7, 8, 12, 16$, respectively, as the same of NeRV evaluation process. The example command is as following:
\begin{verbatim}
python train_nerv.py -e 50 --lower-width 96 --num-blocks 1 --frame_gap 1 
            --embed 1.25_80  --stem_dim_num 512_1  
            --reduction 2  --fc_hw_dim 9_16_26 --expansion 1 
            --single_res --loss Fusion6   --warmup 0. --lr_type cosine  
            --strides 5 3 2 2 2  --conv_type conv -b 1  --lr 0.0005 
            --norm none --act swish --prune_ratio 0.4  
            --eval_only --quant_axis 0 --dataset  LSCD/complex 
            --lscd_level complex --autoresume
            --test_lscd_list dataset_subset/complex.csv  
            --quant_bit_list 4 5 6 7 8 12 16
\end{verbatim}

\end{document}